\documentclass[pra,aps,amsmath,amssymb,amsfonts,twocolumn,nofootinbib,longbibliography,superscriptaddress]{revtex4-1}
\usepackage{amssymb}
\usepackage{bm,mathrsfs}
\usepackage{graphicx}
\usepackage{epsfig}
\usepackage{amsmath,bbm}
\usepackage{amsfonts,amssymb}
\usepackage{times}
\usepackage{verbatim}
\usepackage[sort&compress]{natbib}
\usepackage{amsmath}
\usepackage{bm}
\usepackage[colorlinks,breaklinks,linkcolor=blue,anchorcolor=blue,citecolor=blue,urlcolor=blue,dvipdfm]{hyperref}

\begin{document}

\title{Selective Rydberg pumping via strong dipole blockade}

\author{Xiao-Qiang Shao}
\email{shaoxq644@nenu.edu.cn}
\affiliation{Center for Quantum Sciences and School of Physics, Northeast Normal University, Changchun 130024, China}
\affiliation{Center for Advanced Optoelectronic Functional Materials Research, and Key Laboratory for UV Light-Emitting Materials and Technology
of Ministry of Education, Northeast Normal University, Changchun 130024, China}

\begin{abstract}
The resonant dipole-dipole interaction between highly excited Rydberg levels dominates the interaction of neutral atoms at short distances scaling as $1/r^3$. Here we take advantage of the combined effects of strong dipole-dipole interaction and  multifrequency
 driving fields to propose one type of selective Rydberg pumping mechanism. In the computational basis of two atoms $\{|00\rangle, |01\rangle,|10\rangle,|11\rangle\}$, this mechanism allows $|11\rangle$ to be resonantly pumped upwards to the single-excited Rydberg states while the transitions of the other three states are suppressed. From the perspective of mathematical form, we achieve an analogous F\"{o}ster resonance for ground states of neutral atoms. The performance of this selective Rydberg pumping is evaluated using the definition of fidelity for controlled-$Z$ gate, which manifests a characteristic of robustness to  deviation of interatomic distance, fluctuation of F\"{o}ster resonance defect, and spontaneous emission of double-excited Rydberg states. As applications of this mechanism, we discuss in detail the preparation of the maximally entangled symmetric state for two atoms via ground-state blockade, and  the maximally entangled antisymmetric state via engineered spontaneous emission, within the state-of-the-art experiments, respectively.
\end{abstract}

\maketitle

\section{Introduction}

Rydberg blockade is the most representative phenomenon observed in neutral atom systems \cite{pra063419ref2,PhysRevLett.87.037901,Urban,Gaetan}. There is a strong dipole-dipole-type interaction scaling as $1/r^3$ or van der Waals-type interaction scaling as $1/r^6$ between Rydberg states, which can lead to a sufficiently large energy shift compared to the excitation Rabi frequency for the double-excited Rydberg states, thereby preventing two or more atoms being from excited under the action of resonant driving field. The pioneering work of  Jaksch {\it et al.} has made Rydberg blockade the backbone of quantum information processing based on coherent dynamics \cite{PhysRevLett.102.170502,pra042306ref1,Browaeys_2016,Saffman_2016,PhysRevLett.100.170504,
PhysRevLett.119.160502,PhysRevA.97.032310,PhysRevLett.121.123605,pra012337ref18,PhysRevLett.124.070503,Bai_2020}.
In particular, recent progress with already state-of-the-art neutral atom systems has demonstrated controlled phase gate fidelity $>95\%$ and entanglement state fidelity $>99\%$ including all experimental imperfections on
timescales of hundreds of nanoseconds \cite{PhysRevLett.123.170503,Madjarov}. It should be noted that in the above scenarios, the effect of spontaneous emission of Rydberg states is minimized by resorting to fast Rydberg pulses, whose durations are far less than the finite lifetimes of Rydberg states for realistic atoms.

Remarkably, the spontaneous emission of Rydberg states plays a completely opposite role in the schemes based on dissipative dynamics, which can be exploited as a resource to cooling atoms into a maximally entangled state irrespective of initial states. Physically, the density operator of this maximally entangled state, as the unique steady-state solution to the Markovian master equation, must not only commute with the Hamiltonian operator of system, but also consist of ground states so as not to be affected by spontaneous radiation. Nevertheless, the Rydberg blockade does not apply to this area due to the fact that any Bell basis formed in the ground-state space will be pumped into the excited Rydberg states.

Compared to the Rydberg blockade, the Rydberg antiblockade is more selective for the ground states \cite{PhysRevLett.98.023002,PhysRevLett.104.013001}. For a bipartite system, it permits a resonant
two-photon transition between one computational basis state and the doubly excited Rydberg state as the shifting energy of Rydberg states
is compensated by the two-photon detuning, while leaving other three computational basis states unchanged. This feature has enabled Rydberg antiblockade to become an effective method for dissipative approach to entanglement in neutral atom systems. For example, Carr and Saffman utilized  the angular degrees
of freedom dependent Rydberg antiblockade and atomic spontaneous emission to obtain high fidelity  entanglement and antiferromagnetic
states \cite{pra012319ref19}, and this scheme was then extended to higher dimensional entanglement \cite{PhysRevA.89.052313} and simplified in Ref.~\cite{PhysRevA.92.022328}. With the help of Rydberg cavity quantum electrodynamics \cite{PhysRevA.82.053832,Grankin_2014}, the tripartite Greenberger-Horne-Zeilinger and W states were also capable of being prepared through the dissipative Rydberg pumping \cite{PhysRevA.96.062315,Li:18}.
However, attention needs to be paid to the condition of Rydberg antiblockade. Except for the unnecessary stark shifts caused by second-order perturbation, it usually requires to precisely control the value of the blockade shift and thus are sensitive to the fluctuations of interatomic distance.


In this paper, we aim to engineer an alternative interaction mechanism for neutral atom systems, which is able to replace the traditional second-order dynamics $\sim\Omega^2/\Delta$ in the Rydberg antiblockade with first-order interaction strength $\sim\Omega$, where $\Omega$ means the Rabi frequency of external driving field coupling the ground state and the Rydberg state and $\Delta$ is the single-photon detuning parameter. This idea mainly comes from our precious work about the unconventional Rydberg pumping (URP) \cite{PhysRevA.98.062338}, where the evolution of two atoms initialized in the same ground state is frozen even under driven by laser fields. Instead of
using the van der Waals interaction of Rydberg atoms for URP, here we take advantage of the combined effects of strong dipole-dipole interaction (which is not replaceable) and  multifrequency
 driving fields to propose a selective Rydberg pumping (SRP) mechanism, i.e., considering the four ground states of two atoms $\{|00\rangle, |01\rangle,|10\rangle,|11\rangle\}$, this mechanism allows $|11\rangle$ to be resonantly pumped upwards to the single-excited Rydberg states while the transitions of the other three states are suppressed. Therefore, for atoms in ground states, our scheme can achieve the same effect as the Rydberg antiblockade, but with more advantages. First, the dynamic evolution of system is to first order dependent of the excitation Rabi frequency of driving field, which does not result in unwanted stark shifts. Second, a certain deviation from the desired interatomic distance is allowed since all the double-excited Rydberg states are virtually excited.

The structure of the paper is organized as follows. In
Sec.~\ref{II}, we first illustrate the mechanism of SRP in detail. In
Sec.~\ref{IIInew} we then assess the performance of SRP with currently achievable parameters in neutral atom experiment, using the definition of fidelity for controlled-$Z$ gate. In Sec.~\ref{IV}, we further apply the technology of SRP to realization of ground-state blockade which is exploited to entangle two ground-state atoms with no need for excitation of Rydberg states.
In Sec.~\ref{V}, we organically combine the SRP mechanism with controlled spontaneous emission of Rydberg states to prepare the maximally entangled states in open systems, where the populations of other hyperfine ground states caused by the spontaneous emission can be recycled through laser cooling technology.
In Sec.~\ref{VI}  we finally give a summary of
our manuscript .

\section{basic MECHANISM}\label{II}
\begin{figure}
\centering\scalebox{0.13}{\includegraphics{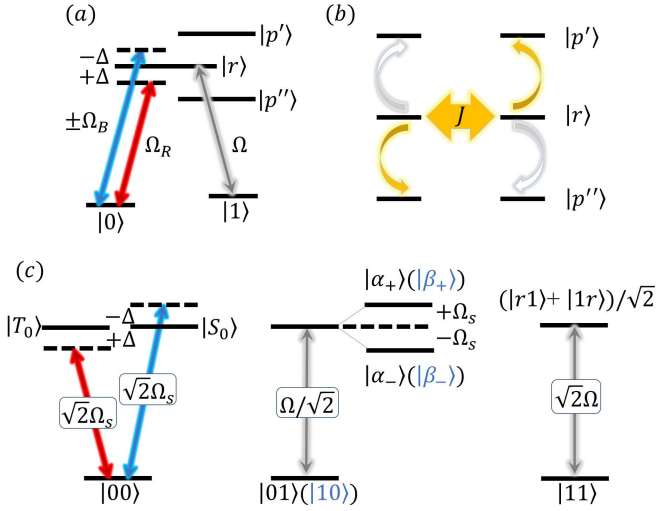}}
\caption{\label{p1}Schematic view of the atomic-level configuration. (a) Each atom is driven by three types of laser fields. The
ground state $|0\rangle$ is dispersively coupled to the Rydberg
state $|r\rangle$ by one laser field with Rabi frequency $\pm\Omega_B$ (which is blue-detuned by $\Delta$) and another laser field with Rabi frequency $\Omega_R$ (which is red-detuned by $\Delta$), simultaneously. While the transition between $|1\rangle$ and $|r\rangle$ is driven by a resonant laser field with Rabi frequency $\Omega$. (b) The dipole-dipole (F\"{o}ster resonance) interaction  of the highly excited Rydberg states. (c) The effective transitions for the ground states, where $|T_0(S_0)\rangle=(|r0\rangle\pm|0r\rangle)/\sqrt{2}$ and $\Omega_s=\Omega_{B(R)}$. In the regime of large detuning  ($\Delta=\sqrt{2}J\gg\Omega_s\gg\Omega$), there is only one resonantly coherent oscillation between state $|11\rangle$ and the single-excited Rydberg states.
}
\end{figure}
The system considered here incorporates two multilevel atoms with the same configuration as adopted in Ref.~\cite{PhysRevLett.87.037901}. Each atom consists of two hyperfine ground states $|0\rangle$ and $|1\rangle$, and three excited Rydberg states $|r\rangle$, $|p'\rangle$, and $|p''\rangle$ as shown in Fig.~\ref{p1}(a). The
ground state $|0\rangle=|5S_{1/2},F=1,m_F=1\rangle$ is dispersively coupled to the Rydberg
state $|r\rangle$ by one laser field with Rabi frequency $\pm\Omega_B$ (``$+$" for atom 1 and ``$-$" for atom 2), blue-detuned by $\Delta$, and another laser field with Rabi frequency $\Omega_R$, red-detuned by $\Delta$, simultaneously. While the transition between $|1\rangle=|5S_{1/2},F=2,m_F=2\rangle$ and $|r\rangle$ is driven by a resonant laser field with Rabi frequency $\Omega$. The resonant dipole-dipole interactions between Rydberg atoms (F\"{o}ster resonance interaction) is
illustrated in Fig.~\ref{p1}(b), which characterizes a hopping transition between a pair of Rydberg states with strength $J$.
In the interaction picture, the Hamiltonian of the system reads
($\hbar$ = 1)
\begin{eqnarray}\label{fH}
H_{I}&=&\sum_{n=1}^{2}\Omega|1\rangle_{n}\langle r|+\{\Omega_{R}e^{-i\Delta t}+\Omega_{B}e^{i[\Delta t+(n-1)\pi]}\}|0\rangle_{n}\langle r|\nonumber\\&&+J|rr\rangle\big(\langle p'p''|+\langle p''p'|\big)+{\rm H.c.},
\end{eqnarray}
where the pair states $|rr\rangle$, $|p'p''\rangle$ and $|p''p'\rangle$ are almost degenerate  \cite{sr}.
In general the strong dipole-dipole interaction between Rydberg atoms can result in a splitting of the excited Rydberg components, thus
this part can be reformulated in the diagonal form $\sqrt{2}J(|E_+\rangle\langle E_+|-|E_-\rangle\langle E_-|)$, where
$|E_{\pm}\rangle=[\sqrt{2}|rr\rangle\pm(|p'p''\rangle+|p''p'\rangle)]/{2}$ are the eigenstates of the Rydberg interaction. For the sake of convenience, we suppose all the Rabi frequencies are real and set $\Omega_B=\Omega_R=\Omega_s$, then the above Hamiltonian, after performing a rotation  with respect to $U=\exp[\sqrt{2}iJt(|E_+\rangle\langle E_+|-|E_-\rangle\langle E_-|)]$,  is rewritten as
\begin{eqnarray}\label{2}
H_{I}&=&H_1+H_2,\\
H_{1}&=&\sqrt{2}\Omega_s\big[|00\rangle\big(\langle T_0|e^{-i\Delta t}+\langle S_0|e^{i\Delta t}\big)\big]\nonumber\\
&&+|01\rangle\big[\frac{\Omega}{\sqrt{2}}(\langle T_0|-\langle S_0|)+2\Omega_s\cos{(\Delta t)}\langle r1|\big]\nonumber\\
&&+|10\rangle\big[\frac{\Omega}{\sqrt{2}}(\langle T_0|+\langle S_0|)-2i\Omega_s\sin{(\Delta t)}\langle 1r|\big]\nonumber\\
&&+\Omega|11\rangle(\langle r1|+\langle 1r|)+{\rm H.c.},\nonumber\\
H_{2}&=&{\Omega_s}|T_0\rangle\big[\langle E_+|e^{-i(\Delta+\sqrt{2}J)t}+\langle E_-|e^{-i(\Delta-\sqrt{2}J)t}\big]\nonumber\\
&&-{\Omega_s}|S_0\rangle\big[\langle E_+|e^{i(\Delta-\sqrt{2}J)t}+\langle E_-|e^{i(\Delta+\sqrt{2}J)t}\big]\nonumber\\
&&+\frac{\Omega}{\sqrt{2}}(|r1\rangle+|1r\rangle)\big(\langle E_+|e^{-\sqrt{2}iJt}+\langle E_-|e^{\sqrt{2}iJt}\big)\nonumber\\&&
+{\rm H.c.},\nonumber
\end{eqnarray}
where $|T_0(S_0)\rangle=(|r0\rangle\pm|0r\rangle)/\sqrt{2}$, $H_1$ describes the interaction between ground states and single-excited Rydberg states, and $H_2$ bridges the transitions of single-excited and two-excited Rydberg states. In
the regime of the large detuning limits $\Delta=\sqrt{2}J$ and $\Delta\gg \{\Omega_s,\Omega\}$, the terms oscillating with high frequencies $\{\pm\Delta,\pm\sqrt{2}J,\pm(\Delta+\sqrt{2}J) \}$ in Eq.~(\ref{2}) can be safely disregarded, then we have a concise form as
\begin{eqnarray}\label{3}
H'_{I}&=&\Omega|11\rangle(\langle r1|+\langle 1r|)+\frac{\Omega}{\sqrt{2}}|01\rangle(\langle T_0|-\langle S_0|)\nonumber\\&&+\frac{\Omega}{\sqrt{2}}|10\rangle(\langle T_0|+\langle S_0|)+{\Omega_s}|T_0\rangle\langle E_-|\nonumber\\
&&-{\Omega_s}|S_0\rangle\langle E_+|+{\rm H.c.}.
\end{eqnarray}
\begin{figure*}
\centering\scalebox{0.5}{\includegraphics{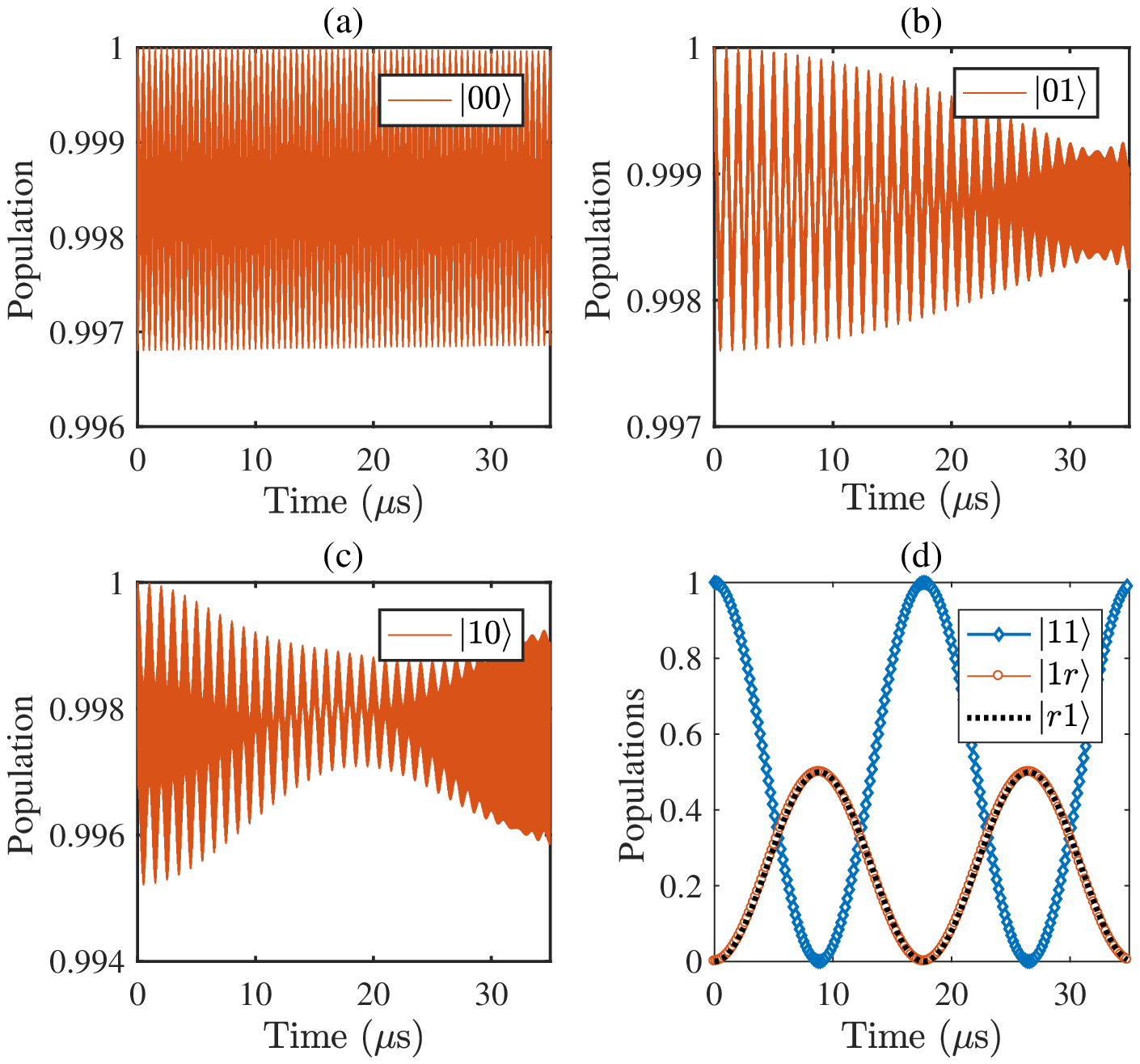}}
\centering\scalebox{0.5}{\includegraphics{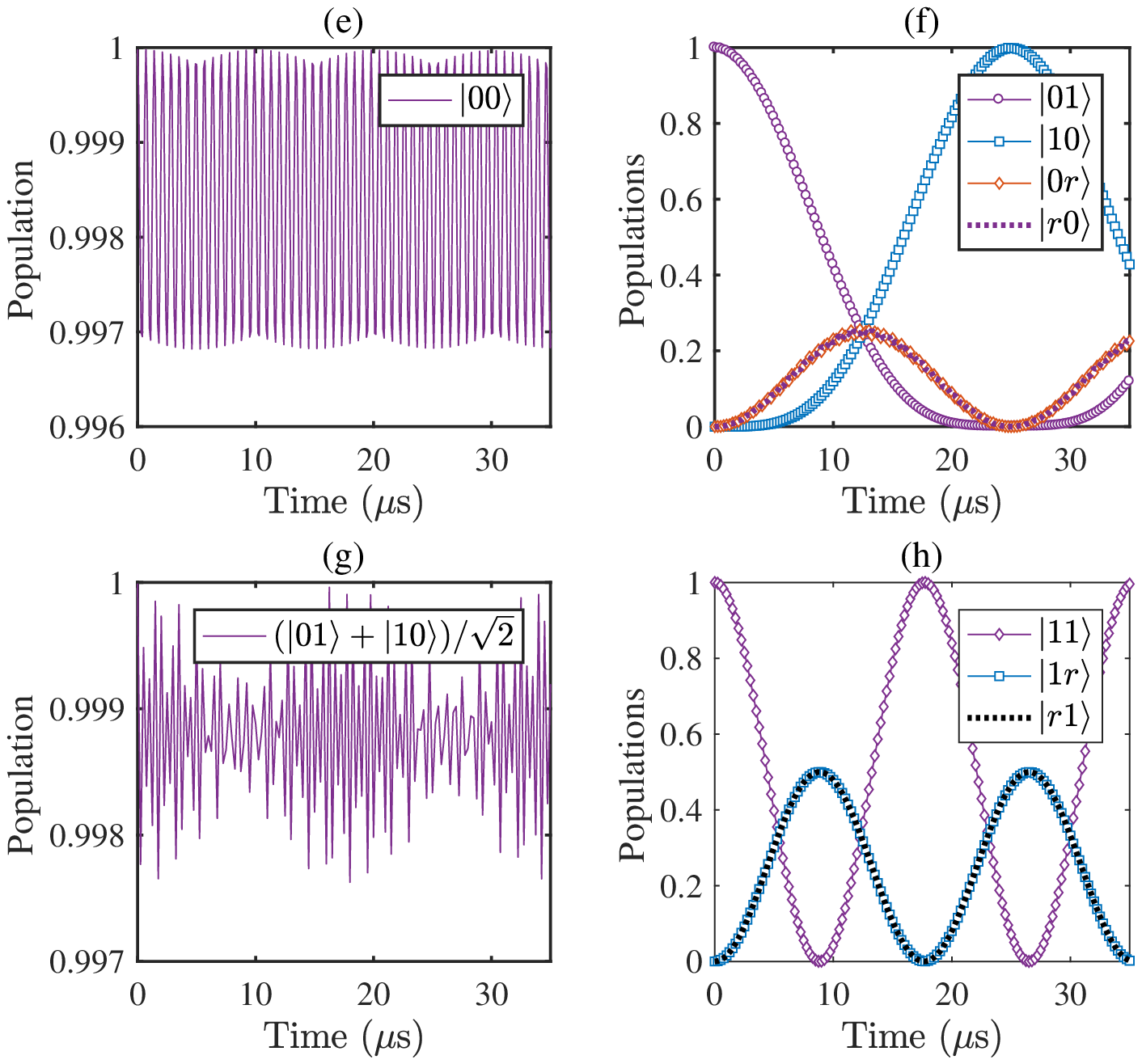}}
\caption{\label{p2} Left panel: The temporal evolution of populations for different initial states \{$|00\rangle$, $|01\rangle$, $|10\rangle$, $|11\rangle$\} corresponding to (a)-(d) governed by the full Hamiltonian Eq.~(\ref{fH}). The states $|00\rangle$, $|01\rangle$, and $|10\rangle$ stay in their initial states and the state $|11\rangle$ is resonantly coupled to the single-excited Rydberg state $(|r1\rangle+|1r\rangle)/\sqrt{2}$. The parameters are chosen
as $\Omega/2\pi=0.02~$MHz, $\Omega_{s}/2\pi=1~$MHz, and $J/2\pi=50~$MHz. Right panel: The corresponding numerical simulations under the van der Waals interaction. The initial state is $|00\rangle$ in (e), $|01\rangle$ in (f), $(|01\rangle+|10\rangle)/\sqrt{2}$ in (g), and $|11\rangle$ in (h), and there is a quantum state transfer process between $|01\rangle$ and $|10\rangle$ mediated by $(|0r\rangle-|r0\rangle)/\sqrt{2}$ as shown in (f) and (g). }
\end{figure*}

This Hamiltonian can be further simplified
by considering the limitation of $\Omega_s\gg\Omega$. Analogous to the process of dipole blockade, the strong Rabi coupling $\Omega_s$ leads to large shifting energies of states $|T_0\rangle$ and $|S_0\rangle$, which are big enough to block the jumps from $|01\rangle$ ($|10\rangle$) towards to the excited states driven by the field of Rabi frequency $\Omega$. Most importantly, there is no extra stark shifts or detuning-induced Raman resonance, because these transition paths mediated by the independent channels $|\alpha_{\pm}\rangle$ and $|\beta_{\pm}\rangle$ interfere destructively, where
\begin{eqnarray}
|\alpha_\pm\rangle&=&\frac{1}{2}[(|T_0\rangle-|S_0\rangle)\pm(|E_+\rangle+|E_-\rangle)],\nonumber\\
|\beta_\pm\rangle&=&\frac{1}{2}[(|T_0\rangle+|S_0\rangle)\mp(|E_+\rangle-|E_-\rangle)],\nonumber
\end{eqnarray}
are the corresponding eigenstates of the part governed by $\Omega_s$. Therefore the Hamiltonian of our current model reduces to an effective form
\begin{equation}\label{4}
H_{\rm eff}=\Omega|11\rangle(\langle r1|+\langle 1r|)+{\rm H.c.}.
\end{equation}
Now we finish the mechanism of SRP. The effective transitions for all ground states are also displayed in Fig.~{\ref{p1}}(c) in order to offer a better physical picture.
Note that there are many interesting features implied in Eq.~(\ref{4}). From the perspective of mathematical form, it describes an analogous F\"{o}ster resonance, where the ground states $|11\rangle$ are coupled to other two states resonantly. So one potential application of our SRP
is to simulate the F\"{o}ster resonance-related phenomena with ground states of neutral atoms.
In addition, the first-order Rabi coupling is easier to achieve experimentally than the second-order interaction as required by the Rydberg antiblockade, the SRP mechanism may greatly simplify the Rydberg-antiblockade-based schemes, while reducing decay of doubly excited Rydberg states.

\section{performance of the SRP}\label{IIInew}
\subsection{Dipole-dipole interaction versus van der Waals  interaction}
According to the works of Browaeys {\it et al.}~{\cite{sr,PhysRevA.92.020701}}, the F\"{o}ster resonance of excited Rydberg states is reached using $|p'\rangle=|61P_{1/2},m_J=1/2\rangle$, $|r\rangle=|59D_{3/2},m_J=3/2\rangle$, and $|p''\rangle=|57F_{5/2},m_J=5/2\rangle$ of two $^{87}$Rb atoms in the presence of an electric field  and  the measured $C_3/2\pi=2.39\pm0.03~$GHz~$\mu {\rm m}^3$ which is close to the theoretical value $C_3/2\pi\simeq2.54~$GHz$~\mu {\rm m}^3$. This allows the dipole interaction strength $J=C_3/R^3$ to be continuously varied between $2\pi\times2.39$~MHz and $2\pi\times152.96$~MHz corresponding to the distance between atoms are adjusted from $10~\mu {\rm m}$ to $2.5~\mu {\rm m}$. The Rabi coupling of the ground states and the Rydberg states is accomplished by a tunable two-photon process that can be up to $2\pi\times5$~MHz in Ref.~\cite{PhysRevA.92.020701}. So the condition for realization of the SRP ($\Delta=\sqrt{2}J\gg\Omega_s\gg\Omega$) is easy to access by choosing $\Omega/2\pi=0.02~$MHz, $\Omega_{s}/2\pi=1~$MHz, and $J/2\pi=50~$MHz. In the left panel of Fig.~\ref{p2}, we depict the temporal evolution of all ground states obtained from the full Hamiltonian of Eq.~(\ref{fH}). It can be seen that this result has an excellent agreement with our prediction. The states $|00\rangle$, $|01\rangle$, and $|10\rangle$ always keep in their initial states with fidelities higher than 99.5\% during the coherent-oscillation process between states $|11\rangle$ and $(|r1\rangle+|1r\rangle)/\sqrt{2}$.

As we mentioned earlier, the dipole-dipole interaction is the key ingredient to implement the SRP, which cannot be substituted by the van der Waals-type interaction. In the right panel of Fig.~\ref{p2} we also analyze the dynamical behavior of each ground state based on the Hamiltonian {$H_{I}=\sum_{n=1}^{2}\Omega|1\rangle_{n}\langle r|+\Omega_{s}e^{i\Delta t}|0\rangle_{n}\langle r|+{\rm H.c.}+U_{\rm vdw}|rr\rangle\langle rr|$} with $U_{\rm vdw}=J=2\pi\times50~$MHz as a comparison. In this case, the energy of the excited Rydberg states will only shift without splitting, which results in an undesired resonant transition between states $(|01\rangle-|10\rangle)/\sqrt{2}$ and $(|0r\rangle-|r0\rangle)/\sqrt{2}$ and thereby destroys the essence of SRP mechanism, as shown in Figs.~\ref{p2}(f) and ~\ref{p2}(g). This is the reason why the current SRP mechanism cannot be achieved by exploiting the van der Waals interaction of neutral atoms.

\subsection{Fluctuations of relevant parameters}
The mechanism of SRP itself defines a kind of quantum logic operation, {\it i.e.},
$|00\rangle\rightarrow|00\rangle$, $|01\rangle\rightarrow|01\rangle$, $|10\rangle\rightarrow|10\rangle$, and $|11\rangle\rightarrow\cos(\sqrt{2}\Omega t)|11\rangle-i\sin(\sqrt{2}\Omega t)(|r1\rangle+|1r\rangle)/\sqrt{2}$, from which the two-qubit controlled-$Z$ gate is readily implemented after $t_g=\pi/(\sqrt{2}\Omega)$. Therefore in order to evaluate more accurately the performance of the SRP mechanism, we use the fidelity of the logic gate defined by \cite{PhysRevA.101.030301}
\begin{equation}\label{srpgate}
F(t_g)=\frac{1}{16}|{\rm Tr}[U^{\dag}(t_g)U_{cz}]|^2,
\end{equation}
where $U_{cz}$ is the ideal controlled-$Z$ gate in the computational basis and $U(t_g)$ is the result of the numerical simulation obtained from our system.

\begin{figure}
\centering\scalebox{0.5}{\includegraphics{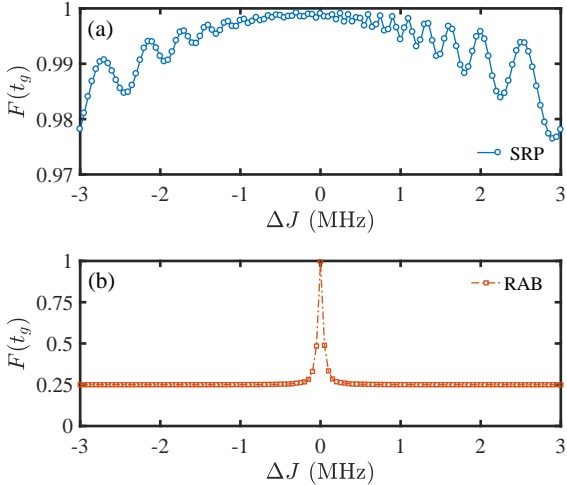}}
\caption{\label{pmis} (a) The effect of deviation $\Delta J$ on the fidelity of the controlled-$Z$ gate based on SRP mechanism. Other parameters:  $\Omega/2\pi=0.02~$MHz, $\Omega_{s}/2\pi=1~$MHz, $\Delta/\sqrt{2}=2\pi\times50$~MHz, $J=2\pi\times(50+\Delta J)$~MHz. (b) The effect of $\Delta J$ on the controlled-$Z$ gate based on Rydberg antiblockade mechanism. The parameters are set as  $\Omega_{s}/2\pi=2^{-1/4}~$MHz $\simeq0.84~$MHz, $\Delta/2\pi=25\sqrt{2}$~MHz $\simeq35.35$~MHz, and $J=2\pi\times(50+\Delta J)$~MHz in order guarantee the same gating time as in the SRP-based scheme.}
\end{figure}

 In the process of deriving the SRP mechanism, we have assumed $\Delta=\sqrt{2}J$. Nevertheless, it is difficult to control the interaction energy of Rydberg states to strictly meet the above condition in experiments. To assess the effect of deviation from the desired dipole-dipole interaction, we suppose $J=2\pi\times(50+\Delta J)$~MHz with the premise that $\Delta/\sqrt{2}=2\pi\times50$~MHz, and plot the fidelity of the controlled-$Z$ gate versus the deviation $\Delta J$ in Fig.~\ref{pmis}(a). It is shown that the current SRP mechanism is insensitive to the fluctuation of the distance between two atoms, because the fidelity of gate remains above 99\% in the continuous range of distance from 3.589~$\mu$m ($\Delta J=1.7~$MHz) to 3.685~$\mu$m ($\Delta J=-2.25~$MHz).
This result is in deep contrast with the rigorous condition of the Rydberg antiblockade effect as shown in Fig.~\ref{pmis}(b), where a minor change on the interatomic distance will greatly destroy the desired dynamics of system \cite{PhysRevA.101.012347,su2020rydberg}.

\begin{figure}
\centering\scalebox{0.5}{\includegraphics{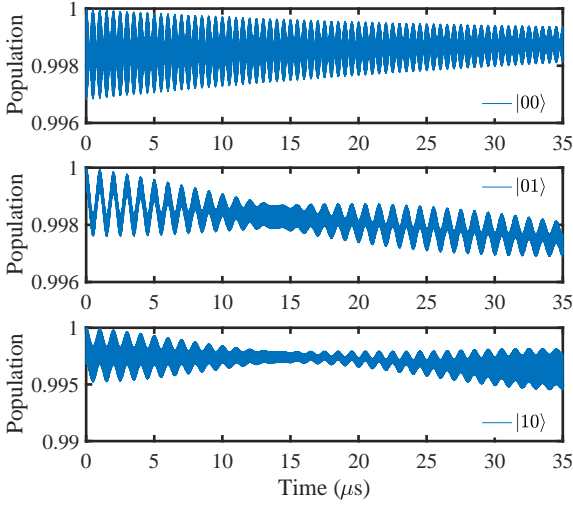}}
\caption{\label{p8} The temporal evolution of populations for states $|00\rangle$, $|01\rangle$,
and $|10\rangle$ in the presence of F\"{o}ster defect $\delta/2\pi$ = 8.5~MHz. Other parameters are the same as in Fig.~\ref{p2}.}
\end{figure}

In the presence of F\"oster defect, the dipole-dipole coupling between two Rydberg states $|rr\rangle$ and $(|p'p''\rangle+|p''p'\rangle)/\sqrt{2}$ in Eq.~(\ref{fH}) is modified by
\begin{equation}\label{def}
H_{dd}=\left(
                                   \begin{array}{cc}
                                     0 & \sqrt{2}J \\
                                    \sqrt{2}J & \delta \\
                                   \end{array}
                                 \right),
\end{equation}
where $\delta$ is the F\"oster defect measuring the detuning of the above two states, and its value is only $2\pi\times8.5~$MHz in the absence of an electric field \cite{sr}. This F\"oster defect alters the eigenvalues of the dipole-dipole interaction and then bring about a deviation $\epsilon=|\sqrt{2}J-(\delta+\sqrt{8J^2+\delta^2})/2|$ of the condition for realization of SRP, which is approximately expanded to $\epsilon\simeq\delta/2+\sqrt{2}\delta^2/16J$ for a small ratio $\delta/J$. In Fig.~\ref{p8} we numerically simulate the temporal evolution of populations for states $|00\rangle$, $|01\rangle$, and $|10\rangle$ in the presence of F\"oster defect $\delta/2\pi=8.5~$MHz, which shows that our SRP still works because the quantum states are well suppressed to their initial states, although the corresponding values are slightly lower than that in Fig.~\ref{p2}. In this sense we can claim that our scheme is robust against the fluctuation of F\"oster defect.

\subsection{Influence of spontaneous emission of Rydberg states}
When the spontaneous emission of the Rydberg states is taken into account, the Markovian master equation of system can be modeled in Lindblad form
\begin{eqnarray}\label{masterl}
\dot\rho&=&-i[H_I,\rho]+\sum_{n=1}^2\sum_{m=r,p',p''}\bigg\{\sum_{j=0}^1\gamma_m^j{\cal D}[|j\rangle_n\langle m|]\nonumber\\&&+\sum_{k=a_1}^{a_n}\gamma_m^k{\cal D}[|k\rangle_n\langle m|]\bigg\},
\end{eqnarray}
where ${\cal D}[|y\rangle\langle x|]=[|y\rangle\langle x|\rho|x\rangle\langle y|-1/2(|x\rangle\langle x|\rho+\rho|x\rangle\langle x|)]$ and $\{|a_1\rangle\dots|a_n\rangle\}$ denotes the subspace consists of external leakage levels out of $\{|0\rangle, |1\rangle\}$ \cite{PhysRevLett.123.170503,Madjarov}. $\gamma_x^y$ is the branching ratio of the spontaneous decay rate from state $|x\rangle$ to $|y\rangle$, which satisfies $\gamma_m=(\sum_{j=0}^1\gamma_m^j+\sum_{k=a_1}^{a_n}\gamma_m^k)=1/\tau_m$. The effective lifetimes of Rydberg states $|p'\rangle=|61P_{1/2},m_J=1/2\rangle$, $|r\rangle=|59D_{3/2},m_J=3/2\rangle$, and $|p''\rangle=|57F_{5/2},m_J=5/2\rangle$ can be inferred from the data provided in Ref.~\cite{PhysRevA.30.2881,PhysRevA.79.052504} through the relation $\tau\sim n^3$, here $n$ is the principle quantum number of Rydberg states. Then we have $\tau_{p'}\simeq0.48~$ms, $\tau_{r}\simeq0.2~$ms, and $\tau_{p''}\simeq0.13~$ms at $T=0~$K.
\begin{figure}
\centering\scalebox{0.5}{\includegraphics{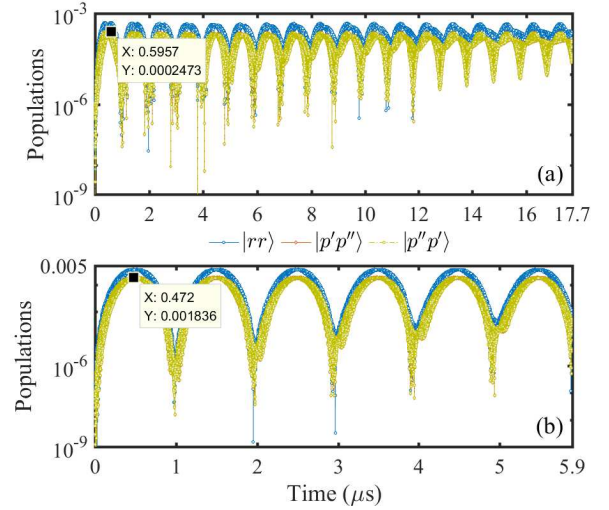}}
\caption{\label{decay} Populations of double-excited Rydberg states $|rr\rangle$, $|p'p''\rangle$, and $|p''p'\rangle$ during the process of SRP with different Rabi frequencies for the resonant driving field. Other parameters: $\Omega_{s}/2\pi=1~$MHz, $J/2\pi=50~$MHz, and $\Omega/2\pi=0.02~$MHz in (a) ($\Omega/2\pi=0.06~$MHz in (b)).}
\end{figure}

As a matter of fact, compared with the effect of spontaneous emission of state $|r\rangle$, the spontaneous emission of states $|p'\rangle$ and $|p''\rangle$ can be ignored  since these two states only appear in pairs and are virtually excited in the process of SRP. In Fig.~\ref{decay}, we use the unitary part of Eq.~(\ref{masterl}) to simulate the  populations of double-excited Rydberg states $|rr\rangle$, $|p'p''\rangle$, and $|p''p'\rangle$ during the implementation of controlled-$Z$ gate from the initial state
$(|00\rangle+|01\rangle+|10\rangle+|11\rangle)/2$.
In Fig.~\ref{decay}(a) we employ the same group of parameters used in Fig.~\ref{p2} to suppress the population of state $|p'p''\rangle (|p''p'\rangle)$ well below $2.5\times10^{-4}$ after the gating time $t\simeq17.7~\mu$s. In Fig.~\ref{decay}(b) we further enhance the Rabi frequency of the resonant driving field to $\Omega/2\pi=0.06~$MHz, which thence shortens the gating time to $t\simeq5.9~\mu$s and can still keep the population of state $|p'p''\rangle (|p''p'\rangle)$ less than $1.9\times10^{-3}$. Consequently, the upper bounds of the effective decay rate for state $|p'\rangle(|p''\rangle)$ can be roughly calculated as $2.5\times10^{-4}/\tau_{p'(p'')}$ and $1.9\times10^{-3}/\tau_{p'(p'')}$ under the above two conditions respectively, which are far less than the decay rate $1/\tau_r$ of state $|r\rangle$.

In addition, the terms of $\sum_{k=a_1}^{a_n}\gamma_m^k{\cal D}[|k\rangle\langle m|]$ in Eq.~(\ref{masterl}) can be substituted by $\gamma_m^{\alpha}{\cal D}[|\alpha\rangle\langle m|]$ for the sake of simplifying calculations, i.e., the effect of the Rydberg states decaying into the multiple non-computational levels
$\{|a_1\rangle\dots|a_n\rangle\}$ is replaceable by only considering the Rydberg states decaying into a single external state $|\alpha\rangle$, provided that the condition $\sum_{k=a_1}^{a_n}\gamma_m^k=\gamma_m^{\alpha}$ is satisfied \cite{PhysRevA.101.062309}. Accordingly the form of Eq.~(\ref{masterl}) can be simplified as
\begin{eqnarray}\label{newmaster}
\dot\rho&=&-i[H_I,\rho]+\sum_{n=1}^2\bigg\{\gamma_r^0{\cal D}[|0\rangle_n\langle r|]+\gamma_r^1{\cal D}[|1\rangle_n\langle r|]\nonumber\\&&+\gamma_r^{\alpha}{\cal D}[|\alpha\rangle_n\langle r|]\bigg\}.
\end{eqnarray}


\begin{table}
\caption{\label{table} Fidelity $F(t_g)$ of the controlled-$Z$ gate with different Rabi frequencies for the resonant driving field, and other parameters are $\Omega_{s}/2\pi=1~$MHz and $J/2\pi=50~$MHz.}
\centering
\scalebox{1}{
\begin{tabular}{p{4cm}l}
\hline\hline
$\Omega/2\pi=0.02~$MHz &$\Omega/2\pi=0.06~$MHz\\\hline
$F^1(t_g)=98.98\%$&$F^1(t_g)=99.48\%$\\
$F^{0.5}(t_g)=98.88\%$&$F^{0.5}(t_g)=99.46\%$\\
$F^{0.2}(t_g)=98.84\%$&$F^{0.2}(t_g)=99.44\%$\\
$F^0(t_g)=98.80\%$&$F^0(t_g)=99.42\%$\\\hline\hline
\end{tabular}}
\end{table}

In Table~\ref{table}, the effect of spontaneous emission of Rydberg states on fidelity of the controlled-$Z$ gate is calculated using Eq.~(\ref{newmaster}), corresponding to different Rabi frequencies for the resonant driving field, where the superscript of $F^{\lambda}(t_g)$ stands for the branching ratio of spontaneous decay rate of Rydberg state $|r\rangle$ into the computational basis $\{|0\rangle, |1\rangle\}$, i.e., $\gamma_r^0+\gamma_r^1=\lambda\gamma_r$, and one can easily check that these values are not affected even the spontaneous emission of state $|p'\rangle(|p''\rangle)$ is considered. It is worth noting that in the above simulation we have postulated $\gamma_r^0=\gamma_r^1=(\gamma_r-\gamma_r^{\lambda})/2$ without loss
of generality, and a selection of other values of ($\gamma_r^0,\gamma_r^1$) does not significantly change the fidelity of gate. Taking the case of \{$\Omega/2\pi=0.06~$MHz, $\lambda=0.5$\} as an example, the minimum (maximum) of fidelity $F^{0.5}(t_g)$ is equal to 99.44\% (99.48\%) under the condition  $\gamma_r^0=0.5\gamma_r$ ($\gamma_r^0=0$).
\begin{figure}
\centering\scalebox{0.5}{\includegraphics{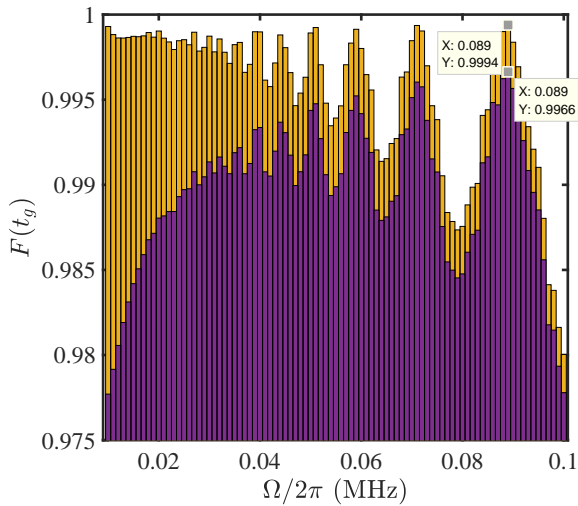}}
\caption{\label{optimal} Fidelity $F(t_g)$ of the controlled-$Z$ gate versus the Rabi frequency for the resonant driving field by fixing other parameters as $\Omega_{s}/2\pi=1~$MHz, and $J/2\pi=50~$MHz. The yellow (higher) bar represents the ideal case and the purple (lower) bar means the worst case that all Rydberg states spontaneously radiate into the external level $|\alpha\rangle$.}
\end{figure}

We also can see from the above analysis that a proper adjustment of the Rabi frequency for the resonant driving field is able to improve the fidelity of quantum gate in the presence of decoherence. Thence we characterize the relationship between $F(t_g)$ and $\Omega$ in Fig.~\ref{optimal} and find an optimal value $\Omega/2\pi=0.089~$MHz within the parameters we consider. This option guarantees a fidelity of 99.94\% at $t\simeq3.97~\mu$s under ideal condition and 99.66\% even all Rydberg states spontaneously decay into the external level $|\alpha\rangle$. There is no doubt that an alternative optimal value of $\Omega$ will appear if other parameters change, which can be determined in a similar way and will not be discussed in more detail.

\section{Entangling Rydberg atoms via ground-state blockade}\label{IV}
\begin{figure}
\centering\scalebox{0.14}{\includegraphics{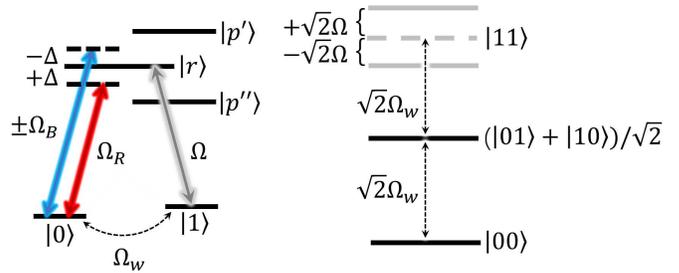}}
\caption{\label{prg} Schematic view of the interactions between atoms and external driving fields for ground-state blockade. Starting from state $|00\rangle$, the population of state $|11\rangle$ is blockaded as $\Omega_w\ll\Omega$. }
\end{figure}
\begin{figure}
\centering\scalebox{0.5}{\includegraphics{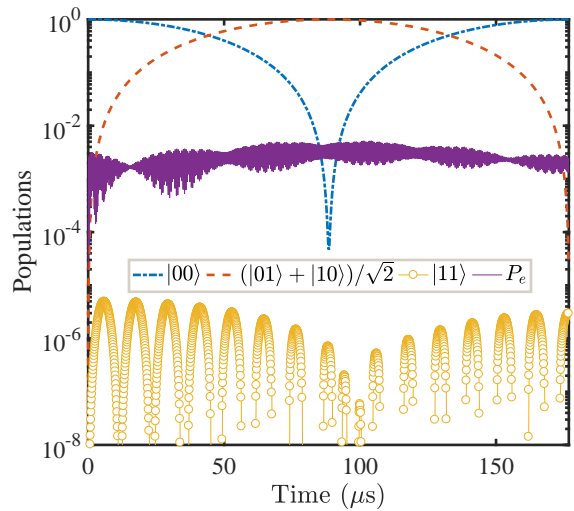}}
\caption{\label{p6} The ground-state blockade induced by the SRP. The Rabi frequency of the laser field that acts on the ground state space is chosen as  $\Omega_w/2\pi=0.002~$MHz, and other parameters are $\Omega/2\pi=0.06~$MHz, $\Omega_{s}/2\pi=2~$MHz, and $J/2\pi=100~$MHz.}
\end{figure}

Using a continuous laser field coupling the ground state to the Rydberg state can form a pair of dressed states. These Rydberg-dressed states reserve a fraction of RRI depending on the probability of Rydberg component, so they
 have a prolonged lifetime as compared to the bare Rydberg state, while maintaining a certain strength of the RRI to induce the blockade effect \cite{PhysRevLett.100.013002,PhysRevA.82.033412}. Therefore the Rydberg dressing technique provides a tunable interaction for quantum control and quantum computing \cite{PhysRevA.87.052314,PhysRevA.85.053615,PhysRevA.87.051602,Balewski_2014,Jau,PhysRevA.96.053417,PhysRevA.95.041801,PhysRevA.97.033414,PhysRevResearch.2.023290}.
By employing the Rydberg-dressed spin-flip blockade, Jau {\it et al.} experimentally produce single-step Bell-state entanglement between two $^{133}{\rm Cs}$ atoms singly trapped in optical tweezers with a fidelity $\geq$ 81(2)\% \cite{Jau}. Nevertheless, the population of the Rydberg state in the dressed states will lead to a probability that the atom will not be recaptured in the trap. When the atom loss events is considered, the entanglement fidelity is reduced to $\sim$ 60\%.

In order to overcome the problem of losing atoms due to the excitation of Rydberg states for preparing Bell-state entanglement, we here realize the  ground-state blockade with current SRP, and  explore the application of the ground-state blockade in entanglement state preparation. Similar to Ref.~\cite{Jau}, we introduce a weak interaction $\Omega_w$ driving the transition between ground states on the basis of Eq.~(\ref{4}), which can be realized directly by a Raman laser field or a microwave field. In this case the effective Hamiltonian of system is
\begin{equation}\label{gb}
H'_{\rm eff}=\sum_{i=1}^{2}\Omega_w|0\rangle_{i}\langle 1|+\Omega|11\rangle(\langle r1|+\langle 1r|)+{\rm H.c.},
\end{equation}
where the energy of state $|11\rangle$ in the ground state space is split by coupling to the single-excited Rydberg state, as shown in Fig.~\ref{prg}. If the Rabi coupling strength $\Omega$ is much larger than the weak driving strength $\Omega_w$, the population of state $|11\rangle$ will be blockaded, and the evolution of atoms is confined to the subspace $\{|00\rangle,|01\rangle,|10\rangle\}$ governed by
\begin{equation}
H_{gb}=\Omega_w|00\rangle(\langle 01|+\langle 10|)+{\rm H.c.}.
\end{equation}

Fig.~\ref{p6} characterizes the evolutions of all ground states beginning with state $|00\rangle$, where we have supposed that all Rydberg states spontaneously decay out of the computational basis states. The weak Raman coupling is selected as $\Omega_w/2\pi=0.002~$MHz, and other parameters are $\Omega/2\pi=0.06~$MHz, $\Omega_{s}/2\pi=2~$MHz, and $J/2\pi=100~$MHz. It clearly reveals the characteristics of ground-state blockade because the population of state $|11\rangle$ is suppressed well below the order of magnitude $10^{-5}$. Meanwhile, a maximally entangled state of two neutral atoms in the form of $|\Psi^+\rangle=(|01\rangle+|10\rangle)/\sqrt{2}$ can be produced with  fidelity $F=\langle\Psi^+|\rho(t)|\Psi^+\rangle=99.66\%$ at $t\simeq88.39~\mu$s through the ground-state blockade effect, as illustrated by the red dashed line of Fig.~\ref{p6}. As for the total probability $P_e={\rm Tr}[\rho(t)(I-\sum_{i,j={0,1,\alpha}}|i\rangle\langle i|\otimes|j\rangle\langle j|)]$ of excitation in Rydberg states, it is always kept below $5.2\times10^{-3}$ (purple solid line) in the preparation of entanglement, which in turn reduces the probability of atom loss efficiently.
\section{Stationary entanglement via engineered spontaneous emission}\label{V}

\begin{figure}
\centering\scalebox{0.5}{\includegraphics{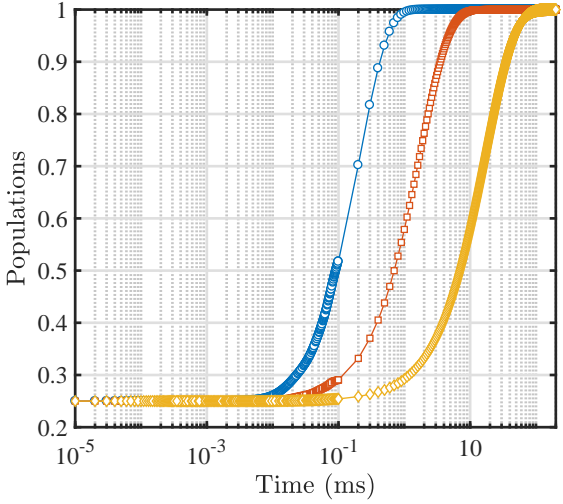}}
\caption{\label{p7} The time evolution of population for the target state governed
by the effective master equation of Eq.~(\ref{master}). The initial
state is a fully mixed state $\rho_0 = \sum_{i,j=0,1}|ij\rangle\langle ij|/4$, and the relevant parameters are set as $\Omega_w/2\pi=0.005~$MHz and $\Omega/2\pi=0.01~$MHz. These results reflect the effect of different values of $\gamma$ on the convergence time of entanglement, where the blue circle denotes $\gamma=0.1~$MHz, the red square represents $\gamma=0.01~$MHz, and the yellow rhombic corresponds to $\gamma=0.001~$MHz. }
\end{figure}
The unitary-dynamics-based protocol requires to exactly tailor the initial state and precisely control the interaction time. In contrast, the reservoir-engineering approaches to entanglement generation is able to release the above
two restrictions by converting the decoherence factor into a resource \cite{pra012319ref19,PhysRevA.92.022328,PhysRevA.89.052313,PhysRevA.95.062339,PhysRevLett.124.070503,PhysRevLett.111.033606,PhysRevA.96.062315,Li:18,Su2020}.
On the basis of the previous model of Fig.~\ref{prg}, we consider the dissipative dynamics of system described by the following effective master equation
\begin{equation}\label{master}
\dot\rho=-i[H'_{\rm eff},\rho]+\sum_{n=1}^2\bigg\{\frac{\gamma}{2}{\cal D}[|0\rangle_n\langle r|]+\frac{\gamma}{2}{\cal D}[|1\rangle_n\langle r|]\bigg\},
\end{equation}
where $H'_{\rm eff}$ is the Hamiltonian of Eq.~(\ref{gb}) and we have assumed that atom decay from the excited Rydberg state $|r\rangle$ into ground states $|0\rangle$ and $|1\rangle$ with the same branching ratio $\gamma/2$, which is useful for proof-of-principle for the underlying mechanism (the real Rb atom case is discussed below). A simple inspection shows that the singlet state $|\Psi^-\rangle=(|01\rangle-|10\rangle)/\sqrt{2}$
 is the unique stationary state solution of Eq.~(\ref{master}).
In Fig.~\ref{p7}, we investigate the influence of different spontaneous emission rates on the convergence time of the maximally entangled state by fixing other parameters as $\Omega_w/2\pi=0.005~$MHz and $\Omega/2\pi=0.01~$MHz from a fully mixed state $\rho_0 = \sum_{i,j=0,1}|ij\rangle\langle ij|/4$. It reveals that a large decay rate can accelerate the convergence time. For example, the choice of $\gamma=0.1~$MHz ensures a fidelity of 99.77\% is achievable at $t=1.2~$ms (blue circle), while a selection of $\gamma=0.001~$MHz delay the time to 100~ms (yellow rhombic).

\begin{figure}
\centering\scalebox{0.5}{\includegraphics{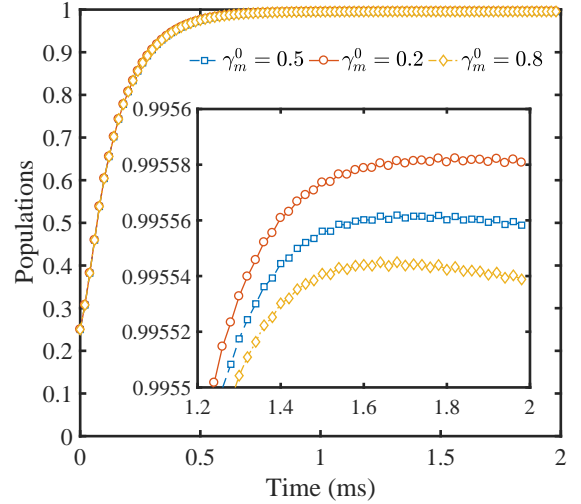}}
\caption{\label{steady} The populations of the target state are simulated
using the full master equation of Eq.~(\ref{master1}) by considering different branching ratios of the spontaneous emission rate $\gamma_m^0$ $(m=r,p',p'')$. The corresponding parameters are set as $\Omega_w/2\pi=0.005~$MHz,  $\Omega/2\pi=0.01~$MHz, $\Omega_s/2\pi=1~$MHz, $J/2\pi=100~$MHz, and $4\Omega_p^2/\Gamma=2\pi\times0.03~$MHz, and  the inset
shows the zoom-in populations from $t=1.2$~ms to $t=2$~ms.}
\end{figure}

We know the effective lifetime of Rb $59D_{3/2}$ Rydberg states is about 0.2~ms ($\gamma=5~$kHz) at $T=0$~K  \cite{PhysRevA.79.052504}, and such a lifetime will lead to a convergence time of $20$~ms for the target state according to the above discussion. Fortunately the decay rate of Rydberg state is adjustable employing the method of engineered spontaneous emission \cite{PhysRevA.98.042310} as shown in Fig~\ref{ps}. By coupling the Rydberg state $|r\rangle=|59D_{3/2},m_J=3/2\rangle$ to a short-lived state $|5P_{3/2},F=2,m_F=2\rangle$ (lifetime $1/\Gamma\simeq$ 25.69~ns) with a weak driving field $\Omega_p$, we may control the Rydberg state $|r\rangle$ to spontaneously decay into the computational basis states $|0\rangle$ and $|1\rangle$ with effective rates $\gamma_{\rm eff}^0=0.6\times4\Omega_p^2/\Gamma$ and $\gamma_{\rm eff}^1=0.4\times4\Omega_p^2/\Gamma$ (see appendix \ref{A} for details). The populations of other hyperfine ground states caused by the spontaneous emission can be recycled via a series of $\sigma^+$ polarized lights which do not disturb the computational basis states (see appendix \ref{B} for details) \cite{pra012319ref19} . After taking all these factors into account, the dynamics of two real Rb atoms can be modeled by the following full master equation
\begin{figure}
\centering\scalebox{0.14}{\includegraphics{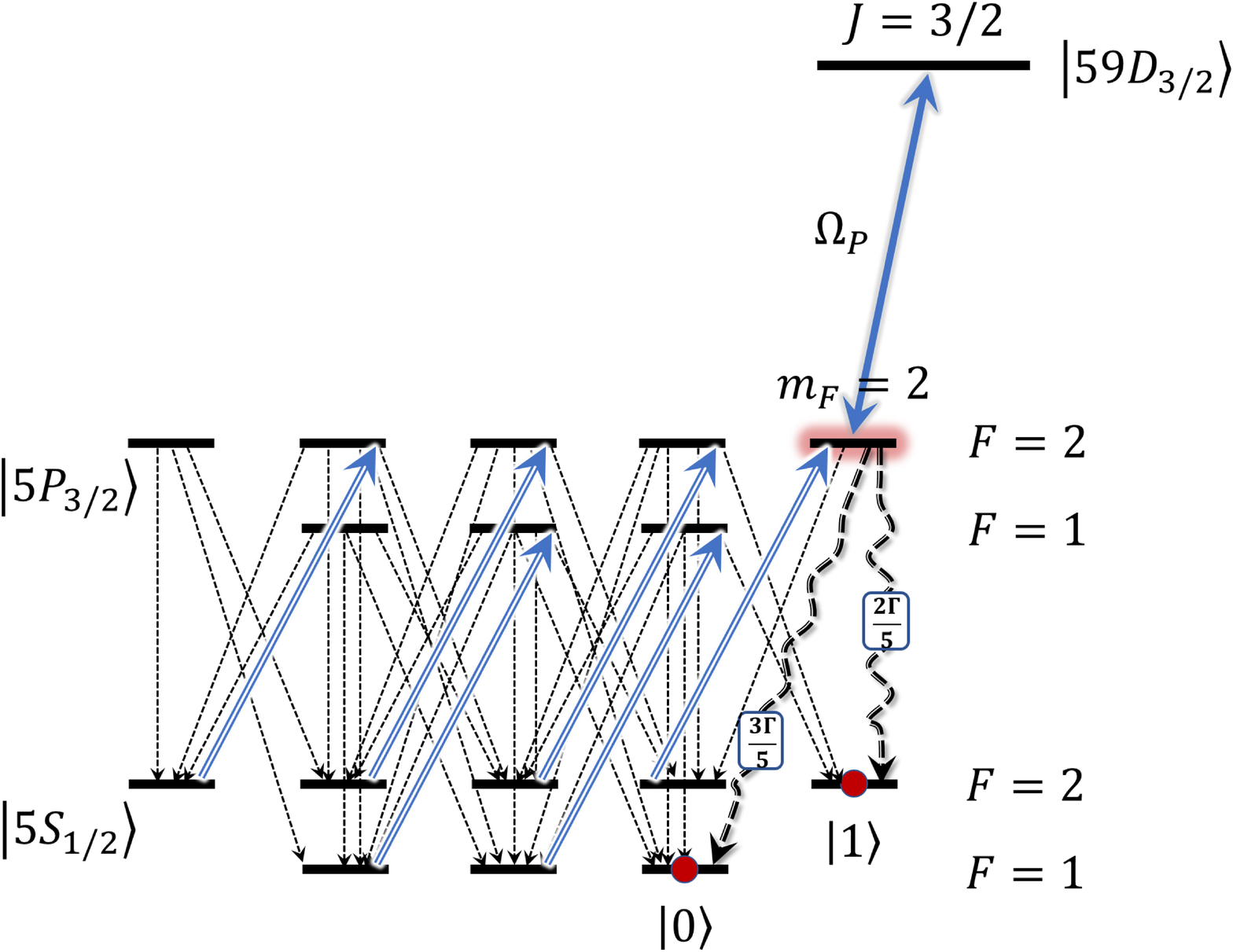}}
\caption{\label{ps} Schematic diagram of the engineered spontaneous emission. A resonantly $\pi$ polarized light (Rabi frequency $\Omega_p$) is used to couple the Rydberg state $|r\rangle=|59D_{3/2}\rangle$  to a short-lived state $|5P_{3/2},F=2,m_F=2\rangle$ (lifetime $1/\Gamma\simeq$ 25.69~ns) which then decays into $|0\rangle=|5S_{1/2},F=1,m_F=1\rangle$ and $|1\rangle=|5S_{1/2},F=2,m_F=2\rangle$ with probabilities $3/5$ and $2/5$, respectively. The final effect is equivalent to the Rydberg state $|r\rangle$ spontaneously decaying into the computational basis states $|0\rangle$ and $|1\rangle$ with effective rates $0.6\times4\Omega_p^2/\Gamma$ and $0.4\times4\Omega_p^2/\Gamma$. The populations of other hyperfine ground states caused by the spontaneous emission can be recycled via a series of $\sigma^+$ polarized lights which do not disturb the computational basis states.}
\end{figure}
\begin{eqnarray}\label{master1}
\dot\rho&=&-i[H_{\rm full},\rho]+\sum_{n=1}^2\bigg\{\gamma_{\rm eff}^0{\cal D}[|0\rangle_n\langle r|]+\gamma_{\rm eff}^1{\cal D}[|1\rangle_n\langle r|]
\nonumber\\&&+\sum_{m=r,p',p''}\sum_{j=0}^1\gamma_m^j{\cal D}[|j\rangle_n\langle m|]\bigg\},
\end{eqnarray}
where $H_{\rm full}=\sum_{i=1}^{2}\Omega_w|0\rangle_{i}\langle 1|+{\rm H.c.}+H_I$. In Fig.~\ref{steady}, we first assume all the Rydberg states share the same branching ratio of the spontaneously decaying into ground states, and then simulate the temporal evolution of the target state with three different branching ratios $\gamma_m^0=0.5$ (blue square), $\gamma_m^0=0.2$ (red circle) and  $\gamma_m^0=0.8$ (yellow rhombic), respectively $(m=r,p',p'')$. Although our hypothesis is somewhat specific, the results can  reflect that the steady population of the target state is basically unaffected by the variation of branching ratio of spontaneous emission rate, as shown in the inset of Fig.~\ref{steady}.

\section{summary}\label{VI}
In summary, we have exhibited how to selectively pump the quantum state of neutral atoms using the current technical means and experimental parameters. This SRP mechanism has an analogous form to the F\"oster resonance interaction, which is robust against the deviation of interatomic distance, the fluctuation of F\"{o}ster resonance defect, and the spontaneous emission of double-excited Rydberg states. Note that all the parameters in our paper are selected as the simplest time-independent form in order to explain our mechanism more clearly. The generalization of our mechanism to the cases of soft temporal modulation and multipartite interaction will be investigated in our future study.
We hope that our work may provide a new prospect
with regard to quantum information processing of neutral
atoms.
\section*{acknowledgment}
The author would like to thank Weibin Li for helpful comments and suggestions. The anonymous reviewers are also thanked for constructive
comments that helped in improving the quality of this paper. This work is supported by National Natural Science Foundation of China (NSFC) under Grants No. 11774047.
\appendix
\section{Derivation of the engineered spontaneous emission}\label{A}
For each atom, the Rydberg state $|r\rangle=|59D_{3/2},m_J=3/2\rangle$ is coupled to a short-lived state $|a\rangle=|5P_{3/2},F=2,m_F=2\rangle$ (lifetime $1/\Gamma$) through an external driving field of Rabi frequency $\Omega_p$, which then rapidly decays into the computational basis state $|0\rangle=|5S_{1/2},F=1,m_F=1\rangle$ and $|1\rangle=|5S_{1/2},F=2,m_F=2\rangle$ with probabilities $3/5$ and $2/5$, respectively \cite{Steck}. Since we here focus  only on the engineered spontaneous emission of $|r\rangle$,  the natural spontaneous emission of the Rydberg state is not taken into account, and the corresponding master equation reads
\begin{equation}\label{motionn}
\dot\rho=-i[H,\rho]+\frac{3\Gamma}{5}{\cal D}[|0\rangle\langle a|]+\frac{2\Gamma}{5}{\cal D}[|1\rangle\langle a|].
\end{equation}
where $H=\Omega_p(|r\rangle\langle a|+|a\rangle\langle r|)$. After expanding the density operator of atom
in the form $\rho(t)=\sum_{\alpha,\beta}\rho_{{\alpha,\beta}}(t)|\alpha\rangle\langle \beta| (\alpha,\beta=r,a,0,1)$ and
substituting it into the master equation of Eq.~(\ref{motionn}), we obtain a
set of coupled equations for the atomic matrix elements:
\begin{equation}\label{motion1}
\dot{\rho}_{aa}=i\Omega_p{\rho}_{ar}-i\Omega_p{\rho}_{ra}-\Gamma{\rho}_{aa},
\end{equation}
\begin{equation}\label{motion4}
\dot{\rho}_{ar}=i\Omega_p{\rho}_{aa}-i\Omega_p{\rho}_{rr}-\frac{\Gamma}{2}{\rho}_{ar},
\end{equation}
\begin{equation}\label{motion5}
\dot{\rho}_{a1}=-i\Omega_p{\rho}_{r1}-\frac{\Gamma}{2}{\rho}_{a1},
\end{equation}
\begin{equation}\label{motion6}
\dot{\rho}_{a0}=-i\Omega_p{\rho}_{r0}-\frac{\Gamma}{2}{\rho}_{a0},
\end{equation}
\begin{equation}\label{motion2}
\dot{\rho}_{rr}=i\Omega_p{\rho}_{ra}-i\Omega_p{\rho}_{ar},
\end{equation}
\begin{equation}\label{motion3}
\dot{\rho}_{00}=\frac{3\Gamma}{5}{\rho}_{aa},\  \  \dot{\rho}_{11}=\frac{2\Gamma}{5}{\rho}_{aa},
\end{equation}
\begin{equation}\label{motion7}
\dot{\rho}_{r1}=-i\Omega_p{\rho}_{a1},\  \  \dot{\rho}_{r0}=-i\Omega_p{\rho}_{a0}.
\end{equation}

In the limit of large decay rate $\Gamma\gg\Omega_p$, the short-lived state $|a\rangle$ can be adiabatically eliminated by assuming $\dot\rho_{aa}=\dot\rho_{ar}=\dot\rho_{a1}=\dot\rho_{a0}=0$, then we have
\begin{equation}
{\rho}_{aa}=\frac{4\Omega_p^2}{\Gamma^2}{\rho}_{rr},\  \
{\rho}_{a1}=-\frac{2i\Omega_p}{\Gamma}{\rho}_{r1},
\end{equation}
\begin{equation}
{\rho}_{ar}=-\frac{2i\Omega_p}{\Gamma}{\rho}_{rr},\ \ {\rho}_{a0}=-\frac{2i\Omega_p}{\Gamma}{\rho}_{r0}.
\end{equation}
Taking advantage of these results, Eqs.~(\ref{motion2})-(\ref{motion7}) can be rewritten as
\begin{equation}
\dot{\rho}_{rr}=-\frac{4\Omega_p^2}{\Gamma}{\rho}_{rr},\  \
\dot{\rho}_{00}=\frac{3}{5}\times\frac{4\Omega_p^2}{\Gamma}{\rho}_{rr},
\end{equation}
\begin{equation}
\dot{\rho}_{11}=\frac{2}{5}\times\frac{4\Omega_p^2}{\Gamma}{\rho}_{rr},\  \
\dot{\rho}_{r0(1)}=-\frac{2\Omega_p^2}{\Gamma}{\rho}_{r0(1)},
\end{equation}
from which we can conclude that the current reduced system is described by an effective master equation as below
\begin{equation}\label{master11}
\dot\rho=\gamma_{\rm eff}^0{\cal D}[|0\rangle\langle r|]+\gamma_{\rm eff}^1{\cal D}[|1\rangle\langle r|],
\end{equation}
where $\gamma_{\rm eff}^0=0.6\times4\Omega_p^2/\Gamma$ and $\gamma_{\rm eff}^1=0.4\times4\Omega_p^2/\Gamma$ are the branching ratios of the engineered spontaneous emission for the Rydberg state $|r\rangle$ due to its coupling to a short-lived state. Thus the engineered decay rate $4\Omega_p^2/\Gamma=2\pi\times0.03~$MHz of Rydberg state $|r\rangle$ in Fig.~\ref{steady} can be obtained by choosing $\Omega_p\simeq1.354~$MHz.

\section{Recycling of the non-computational basis states}\label{B}
The method of recycling the non-computational basis states due to the spontaneous emission of excited states essentially utilizes the laser cooling technology. Fig.~\ref{ps} describes the existence of only one external state $|\alpha\rangle$. On the one hand, this instructive model can clearly reveal the principle of dealing with the population of non-computational basis state using recycling laser, on the other hand, it also can reduce our computational complexity. According to the selection rule for atomic transitions \cite{Steck}, the short-lived state $|5P_{3/2},F=2,m_F=2\rangle$ decays into $|0\rangle=|5S_{1/2},F=1,m_F=1\rangle$, $|1\rangle=|5S_{1/2},F=2,m_F=2\rangle$, and $|a\rangle=|5S_{1/2},F=2,m_F=1\rangle$ with probabilities $1/2$, $1/3$, and $1/6$, respectively. Similar to the process of the engineered spontaneous emission, an introduction of a weak pumping field of Rabi frequency $\Omega_b$ that coupled to the transition between $|\alpha\rangle$ and $|5P_{3/2},F=2,m_F=2\rangle$ will increase the populations of states $|0\rangle$ and $|1\rangle$, while decreasing the population of state $|\alpha\rangle$, which can be quantitatively represented in the Lindblad form $\sum_{n=1}^2\sum_{j=0,1,\alpha}\gamma_{\rm p}^j{\cal D}[|j\rangle_n\langle \alpha|]$, where $\gamma_{\rm p}^0=1/2\times4\Omega_{b}^2/\Gamma$, $\gamma_{\rm p}^1=1/3\times4\Omega_{b}^2/\Gamma$, and $\gamma_{\rm p}^{\alpha}=1/6\times4\Omega_{b}^2/\Gamma$.
\begin{figure}
\centering\scalebox{0.14}{\includegraphics{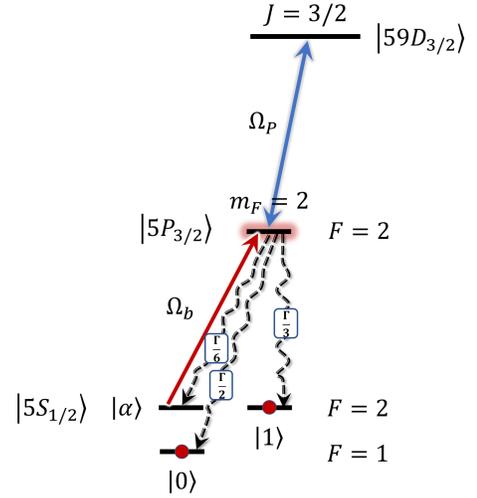}}
\caption{\label{ps} An instructive model for understanding the principle of dealing with the population of non-computational basis state using recycling laser.  The short-lived state $|5P_{3/2},F=2,m_F=2\rangle$ decays into $|0\rangle=|5S_{1/2},F=1,m_F=1\rangle$, $|1\rangle=|5S_{1/2},F=2,m_F=2\rangle$, and $|a\rangle=|5S_{1/2},F=2,m_F=1\rangle$ with probabilities $1/2$, $1/3$, and $1/6$, respectively \cite{Steck}. The combination of the resonant pumping field of Rabi frequency $\Omega_b$ and the decay of the short-lived state contributes additional Lindblad terms $\sum_{n=1}^2\sum_{j=0,1,\alpha}\gamma_{\rm p}^j{\cal D}[|j\rangle_n\langle \alpha|]$ to Eq.~(\ref{master1}).}
\end{figure}
In this case, the master equation of the model reads
\begin{eqnarray}\label{master2}
\dot\rho&=&-i[H_{\rm full},\rho]+\sum_{n=1}^2\bigg\{\sum_{j=0,1,\alpha}\bigg[\gamma_{\rm eff}^j{\cal D}[|j\rangle_n\langle r|]\nonumber\\
&&+
\gamma_{\rm p}^j{\cal D}[|j\rangle_n\langle \alpha|]+\sum_{m=r,p',p''}\gamma_m^j{\cal D}[|j\rangle_n\langle m|]\bigg]\bigg\}.
\end{eqnarray}
\begin{figure}
\centering\scalebox{0.5}{\includegraphics{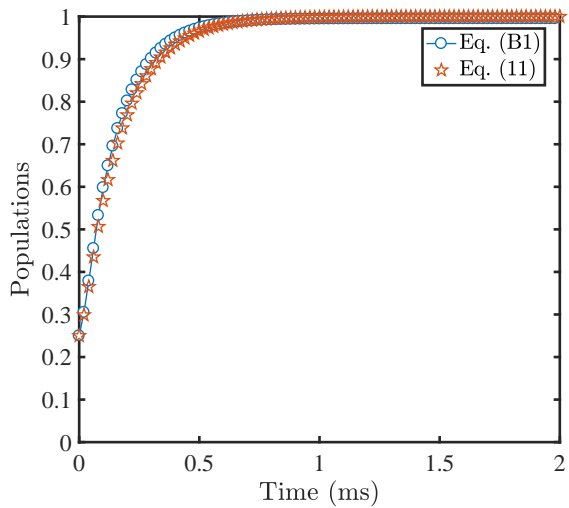}}
\caption{\label{pump} The populations of the target state governed by
 the full master equation of Eq.~(\ref{master2}) and the effective master equation of Eq.~(\ref{master}). The corresponding parameters are set as $\gamma_m^0=\gamma_m^1=0.3\gamma_m$, $4\Omega_{b}^2/\Gamma=2\pi\times0.06~$MHz, and other parameters are the same as in Fig~\ref{steady}.}
\end{figure}
Evidently, the maximally antisymmetric state $|\Psi^-\rangle=(|01\rangle-|10\rangle)/\sqrt{2}$
is still the unique steady state of the system. In Fig.~\ref{pump}, The temporal evolution of population of the target state is plot using Eq.~(\ref{master2}) (blue circle), given that $\gamma_m^0=\gamma_m^1=0.3\gamma_m$, $4\Omega_{b}^2/\Gamma=2\pi\times0.06~$MHz, and other parameters are the same as in Fig~\ref{steady}. The system dynamics evolution given by this result is consistent not only with the dynamic behavior depicted in Fig~\ref{steady}, but also with the dynamic behavior described by the effective master equation of  Eq.~(\ref{master}) (red pentagon). It is not difficult to see that even if there are multiple external states, we can also use laser cooling technology to achieve our goal of preparing the maximally entangled state with dissipation.

\bibliography{newpra.bbl}

\end{document}